\documentclass[12pt]{article}
\usepackage[totalheight = 23cm, totalwidth = 17cm]{geometry}
\usepackage{amssymb,amsmath,amsfonts,amsbsy,graphicx}
\usepackage{color}
\usepackage{psfrag}
\usepackage{cite}
\usepackage{mathrsfs}

\newcommand\be{\begin{equation}}
\newcommand\ee{\end{equation}}
\newcommand\ba{\begin{eqnarray}}
\newcommand\ea{\end{eqnarray}}\newcommand\eq{\begin{equation}}           
\newcommand\en{\end{equation}}

 \newcommand\mchi{m_{\chi}}
 \newcommand\tkd{T_{kd}}

\input epsf

\usepackage{ulem}
\usepackage{amssymb}
\usepackage{amsmath}
\usepackage{bm}
\usepackage{graphicx}
\usepackage{color}
\usepackage{subfigure}
\usepackage{caption}

\def\gsim{\;\rlap{\lower 2.5pt
 \hbox{$\sim$}}\raise 1.5pt\hbox{$>$}\;}
\def\lsim{\;\rlap{\lower 2.5pt
 \hbox{$\sim$}}\raise 1.5pt\hbox{$<$}\;}

\usepackage{graphicx}
\usepackage{url}
\usepackage{amssymb,amsmath}
\usepackage{amsbsy}

\def\Sun{\odot}

\usepackage{graphicx}
\usepackage{url}
\usepackage{amssymb,amsmath}
\usepackage{amsbsy}

\begin{document}
\title{
Late Kinetic Decoupling of Light Magnetic Dipole Dark Matter}
 \author{Paolo Gondolo$^1$ and Kenji Kadota$^2$\\
{ \small $^1$ \it  Department of Physics and Astronomy, University of Utah, Salt Lake City, UT 84112, USA} \\
{ \small $^2$ \it  Center for Theoretical Physics of the Universe, Institute for Basic Science, Daejeon 305-811, Korea} 
}
\date{\vspace{-5ex}}
\maketitle   

\begin{abstract}
We study the kinetic decoupling of light ($\lesssim 10$ GeV) magnetic dipole dark matter (DM). We find that  present bounds from collider, direct DM searches, and structure formation allow magnetic dipole DM to remain in thermal equilibrium with the early universe plasma until as late as the electron-positron annihilation epoch. This late kinetic decoupling leads to a minimal mass for the earliest dark protohalos of thousands of solar masses, in contrast to the conventional weak scale DM scenario where they are of order $10^{-6}$ solar masses.
\end{abstract}

\setcounter{footnote}{0} 
\setcounter{page}{1}\setcounter{section}{0} \setcounter{subsection}{0}
\setcounter{subsubsection}{0}
\section{Introduction}

A characteristic feature of dark matter (DM) is the way the DM couples to Standard Model (SM) particles. Although  weakly interacting massive particles (WIMPs) are typically assumed to couple through a massive force carrier,  long-range force carriers such as the photon remain an interesting possibility. Examples of interactions mediated by the photon are DM with the dipole or higher multipole interactions and milli-charged DM~\cite{sig,dav,turnoff}. Such interactions through a massless or light mediator can be of special interest when the momentum transfer is small, because they exhibit an infrared enhancement in the cross section. They are commonly discussed for light DM (mass $\lesssim 10$ GeV), to which the current DM direct search experiments are not so sensitive \cite{essig2,eee,cora,tashi,kenjisn,celine,boesh,rocky,beam,bur,pos,dubo,mas,bank,ess,semi,light,barg2,heo3,ilidio,gary,kst}. In this paper, we are interested in how long such light DM with a light mediator can stay in kinetic equilibrium with the surrounding hot plasma in the early Universe, and how this may affect the size of the smallest DM protohalos (the first gravitationally bound objects). We are referring to kinetic decoupling, often quoted as thermal decoupling, characterized by the moment at which the rate of DM elastic scattering becomes negligible compared to the Hubble expansion rate. Kinetic decoupling typically occurs much after the chemical decoupling that characterizes the epoch at which the DM annihilation rate becomes comparable to the expansion rate, because the DM annihilation rate is much smaller than the rate at which the DM scatters elastically with the abundant SM particles in the plasma.  
When the DM has a small mass and interacts through a long-range force carrier, the DM tends to stay in kinetic equilibrium longer than the conventional weak scale WIMPs \cite{bring,chen,chen2,hofm2,wimpy,brifirst,bert,kasahara,matias,boe,kam2,pro4,gon}. We therefore pay  particular attention to low kinetic decoupling temperatures and examine the bounds from collider and DM direct search experiments. We present our analysis for the concrete example of a light DM that possesses an anomalous magnetic dipole moment. A sizable magnetic dipole moment arises naturally, for instance, in neutral composite particles such as the neutron. There have been attempts to model composite DM having such dipole moments in a UV-complete theory \cite{dine1,foadi}.  In this paper we do not seek a UV-complete theory, but treat the DM mass and its magnetic dipole moment as free parameters, and we study generic constraints in a model independent manner.

In \S \ref{secformalism} we give our general formalism to estimate the DM kinetic decoupling temperature. We then apply our formalism to the magnetic dipole DM as a concrete example in \S \ref{secmdm}. After a brief review of the protohalo size estimation in \S \ref{secproto}, we explore the experimental bounds on the kinetic decoupling temperature and correspondingly the protohalo size from collider and direct search experiments in \S \ref{secexpbounds}. A discussion and conclusions are given in \S \ref{secdis}. 
 

\section{Dark matter kinetic decoupling}
\label{secformalism}

DM kinetic decoupling occurs when $\gamma(T)$, the rate of momentum transfer between the DM and the thermal plasma of temperature $T$ through the elastic scattering, becomes small compared with the Hubble expansion rate $H(T)$. We estimate the kinetic decoupling temperature $\tkd$ as the solution of $
{\gamma(\tkd)}/{2}=H(\tkd)$ \cite{bert}. To obtain this solution $\tkd$, one can solve the Boltzmann equation without assuming the DM to be a perfect fluid, which can be reduced to the following Fokker-Planck equation, valid for DM mass $\mchi \gg T$ and small momentum transfer,
\ba
\frac{\partial f_{\chi}}{\partial t} - H {\bf p}_{\chi} \cdot \frac{\partial f_\chi}{\partial {\bf p}_\chi} =
\gamma(T)  \,\, \frac{\partial }{\partial {\bf p}_{\chi}}
\cdot \left(
{\bf p}_{\chi} f_{\chi}(1\pm f_\chi)+m_{\chi}T\frac{\partial f_{\chi}}{\partial {\bf p}_{\chi}}
\right).
\label{FP}
\ea
An expression of the momentum relaxation rate for $m_\chi \gg T$ is \cite{gon,kasahara}
\ba
\label{ourgamma}
\gamma(T) &=& \sum_{i} \frac{g_i}{6 m_{\chi} T} \int^{\infty}_{0} \frac{d^3{\bf p}}{(2\pi)^3} \, f_i \, (1\pm f_i) \, \frac{p}{\sqrt{p^2+m_i^2}}  \int^{0}_{-4p^2} dt \, (-t) \, \frac{d\sigma_{\chi+i\to\chi+i}}{dt},
\label{eq:gamma}
\ea
where the sum runs over the plasma particles scattering with the DM, $f_i$ is the phase-space occupation number of a plasma particle species, $g_i$ is its statistical factor (number of degrees of freedom), $m_i$ is the particle mass, ${d\sigma_{\chi+i\to\chi+i}}/{dt}$ is the elastic scattering cross section for $\chi+i\to\chi+i$ differential in the Mandelstam variable $t$, and the sign in $1\pm f_i$ is $+(-)$ for fermions (bosons).
For isotropic distributions $f_i(\omega)$, where $\omega$ is the energy of the plasma particle, we can write $\gamma(T)$ as
\begin{align}
\label{gammaform}
\gamma(T) & =  \sum_{i} \frac{g_i}{384\pi^3 m_\chi^3T} \int_{m_i}^{\infty} d\omega \, \mathscr{F}(p_{\rm cm}) \,
{f}_{i}(\omega) [1\pm {f}_{i}(\omega)]  ,
\end{align}
where 
\begin{align}
\mathscr{F}(p_{\rm cm}) = \int_{-4k^2}^{0} dt \, (-t) \, \overline{|\mathscr{M}|^2} 
\end{align}
with $p_{\rm cm} = k = (\omega^2-m_i^2)^{1/2}$ (the momentum of the plasma particle) and $s=m_\chi^2+m_i^2+2m_\chi\omega$ (these expressions are valid in the heavy-particle limit $m_\chi \gg T,m_i$).
The overline on the invariant scattering amplitude $\mathscr{M}$ indicates the usual sum over final polarizations and average over initial polarizations.  

We also note that the quantity $\mathscr{F}$ is related to the transport cross section
\begin{align}
\label{tra}
\sigma_T = \int d\Omega \, (1-\cos\theta) \, \frac{d\sigma}{d\Omega} ,
\end{align}
where $\theta$ is the scattering angle in the center of mass frame. This follows from the relations
\begin{align}
t = - 2 p_{\rm cm}^2 (1-\cos\theta)
\end{align}
and
\begin{align}
\label{relaform}
\frac{d\sigma}{dt} = \frac{\overline{|\mathscr{M}|^2}}{64\pi s p_{\rm cm}^2} ,
\end{align}
from which
\begin{align}
\mathscr{F}(p_{\rm cm}) 
& = 128 \, \pi \, p_{\rm cm}^4 \, s \, \sigma_T. \label{tcross}
\end{align}
In the heavy-particle limit $m_\chi \gg T$, one has $p_{\rm cm}^4 s \simeq k^4 m_\chi^2 $, and 
\ba
\gamma  = \frac{g_i}{3\pi^2 m_\chi T}  \int_{m_i}^{\infty} d\omega \, k^4  \, \sigma_T \,
f_{i}(\omega) [1\pm f_i(\omega)].
\ea
We apply our formalism to magnetic dipole dark matter in the following section.

\section{Magnetic dipole dark matter}
\label{secmdm}
We consider a spin-1/2 DM particle $\chi$ coupled to photons through a magnetic moment interaction given by the gauge-invariant dimension-five Lagrangian
\begin{align}
\mathscr{L} = -\frac{i}{2} \mu_{\chi} \, \overline{\chi} \, \sigma_{\mu\nu} \! \chi F^{\mu\nu} .
\end{align}
Here $\mu_{\chi}$ is the DM magnetic dipole moment, $F_{\mu\nu}$ is the electromagnetic field, and $\sigma_{\mu\nu}=\frac{i}{2}(\gamma_{\mu} \gamma_{\nu}  - \gamma_{\nu}  \gamma_{\mu} )$. The tensor current $\bar{\chi} \sigma_{\mu \nu} \chi$ vanishes for a Majorana particle, thus the particle we consider is a Dirac fermion. We assume CP invariance and do not hence consider electric dipole operators.
In our quantitative discussions, we give the magnetic dipole moment in units of the nuclear magneton $\mu_N=e/2 m_p$ ($=0.105~e$~fm). From a purely dimensional point of view, one may expect the dark matter magnetic dipole moment to be at most of order $\mu_{\chi}/ \mu_N \sim m_p/m_{\chi}$. Much larger magnetic dipole moments $\mu_\chi$ may hint to a complex nature of the DM particles.

The momentum relaxation rate $\gamma(T)$ depends on the elastic $\chi+i\to\chi+i$ cross section, where $i$ is a charged particle in the plasma. The unpolarized matrix element squared for the interaction of a magnetic-dipole DM particle with a particle of mass $m_i$ and charge $q_ie$ through a photon exchange is
\begin{align}
\overline{\left| \mathscr{M} \right|^2}  = 16 \pi \alpha q_i^2 \mu_\chi^2 \, \frac{-t}{(t-\omega_P^2)^2} \left[ m_\chi^4 -2 m_\chi^2 (s+t+m_i^2) + (s-m_i^2)(s+t-m_i^2) \right],
\end{align}
where $\alpha=e^2/4\pi$ is the electromagnetic fine structure constant, $s$ and $t$ are the conventional Mandelstam variables, and $\omega_P$ is the plasma frequency \cite{kal}. The plasma frequency for a single-species plasma at temperature $T\gg m_i$ is
$ \omega_P = \sqrt{{4\pi \alpha}/{9}} T$, while for
for $T\ll m_i$ it is $\omega_P = \sqrt{{4\pi \alpha n_i}/{m_i}}$; the plasma frequency of the multi-species plasma is the largest of the plasma frequencies of each species. 
 In the heavy-particle limit, $s=m_\chi^2+m_i^2+2m_\chi \omega$, and, neglecting $t$ with respect to $m_\chi^2$,
\begin{align}
  \overline{\left| \mathscr{M} \right|^2}  = 16 \pi \alpha q_i^2 \mu_\chi^2 m_\chi^2 \frac{t(t-4k^2)}{(t-\omega_P^2)^2}
  \end{align}
where $k$ are the momentum of the scattering species. We hence have
\begin{align}
\mathscr{F}(k) = \int_{-4k^2}^{0} dt \, (-t)  \overline{\left| \mathscr{M} \right|^2}  = 384 \pi  \alpha q_i^2 \mu_\chi^2 m_\chi^2 k^4 \,F_{\rm screen}\!\left(\frac{\omega_P}{2k}\right),
\end{align}
where
\begin{align}
F_{\rm screen}(x) = \frac{3+x^2-6x^4+2x^2(-2+x^2+3x^4)\ln(1+x^{-2})}{3(1+x^2)}
\end{align}
is a screening factor with $F_{\rm screen}(0)=1$. 
We then estimate the relaxation rate $\gamma(T)$ from Eq. (\ref{gammaform}) for a relativistic plasma particle.
For the scattering of DM with a non-relativistic plasma particle, we can simplify the estimation of the momentum transfer $\gamma(T)$ as \cite{cora}
\ba
\gamma(T) \approx \frac{2.1 \rho_i \sigma_T}{m_{\chi}+m_i} \left( \frac{T_i}{m_i}+\frac{T_{\chi}}{m_{\chi}} \right)^{1/2},
\ea
where $\rho_i$ is the density of the plasma species and $\sigma_T$ is the transport cross section given in Eq. (\ref{tra}).

We numerically solve the condition $\gamma(T_{kd})/2=H(T_{kd})$ to estimate the kinetic decoupling temperature $\tkd$.  In this paper, we focus on DM masses $m_\chi \lesssim 10$~GeV and correspondingly small kinetic decoupling temperatures. We hence only consider $\tkd <100$ MeV. At $500~$keV~$\lesssim \tkd \lesssim 100$~MeV, the dominant charged plasma particles are the light charged leptons ($q_i=\pm 1$). After the electron-positron annihilation epoch ($\tkd \lesssim 500$~keV), the electron abundance decreases significantly. For such low temperatures, we consider DM scattering off electrons and protons since scattering off protons can dominate the DM momentum transfer when the proton and electron abundances are comparable. In our numerical analysis, the chemical potentials for the electrons and the protons are taken care of by demanding charge neutrality $n_e = n_p$. 

\begin{figure}
\begin{center}    
\epsfxsize = 0.1\textwidth
\includegraphics{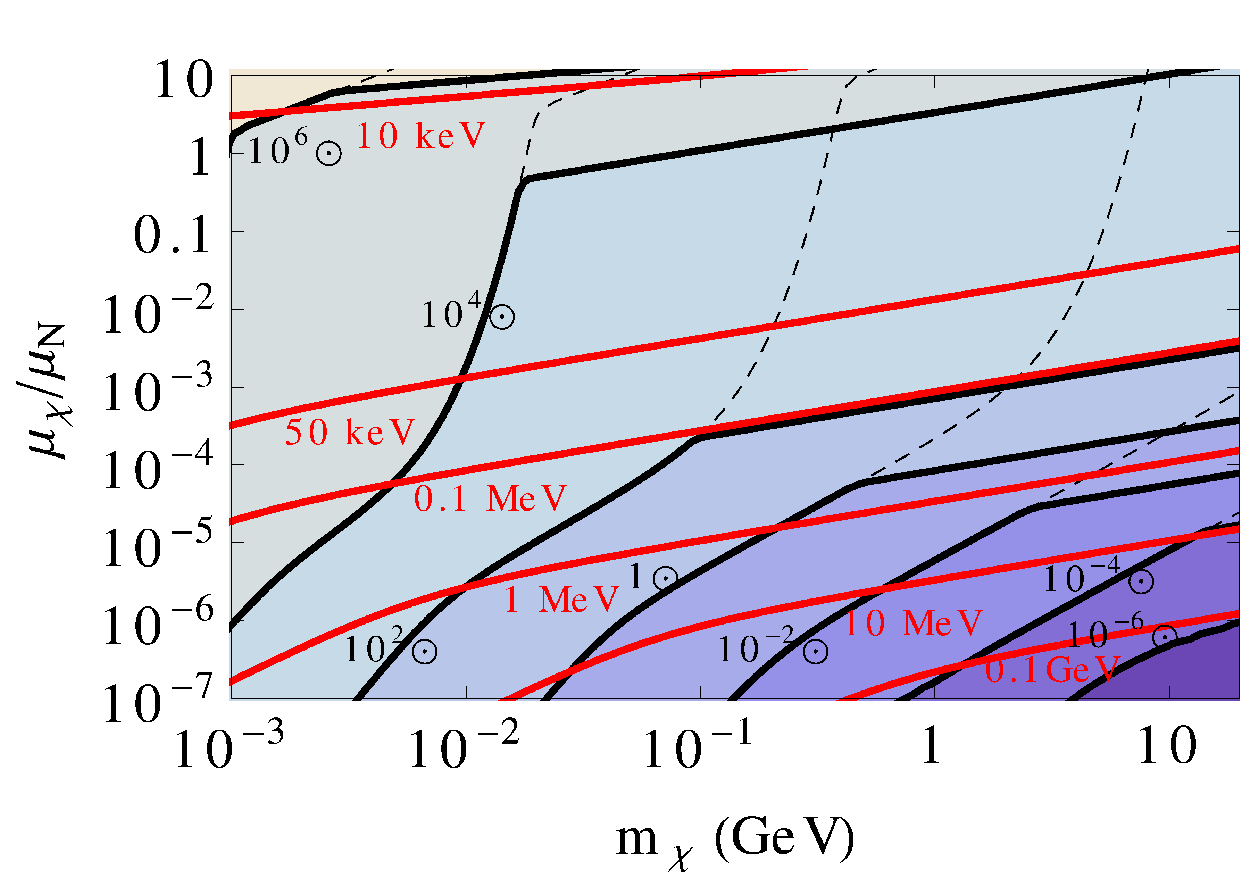}
\end{center}  
\caption{The black solid curves show the smallest protohalo mass (larger of $M_{fs}$ and $M_{kd}$) for magnetic dipole DM. The black dotted curves show $M_{fs}$ (the mass inside the free streaming scale) in the parameter region where $M_{fs}<M_{kd}$ (the mass inside the horizon scale at the time of DM kinetic decoupling). The red contours represent the DM kinetic decoupling temperature. The protohalo size is given in units of solar masses $M_\Sun$, and the DM magnetic moment $\mu_{\chi}$ is in units of nuclear magnetons $\mu_N$ ($\sim 0.1~e$ fm).}
\label{massSept25temp}
\end{figure}

Our results are shown as contour plots of $T_{kd}$ as a function of $m_{\chi}$ and $\mu_{\chi}$ in Fig. \ref{massSept25temp} (red solid lines). A large gap in the $T_{kd}$ contours is observed in correspondence to the electron-positron annihilation epoch. Long after the electron-positron annihilation epoch ($T\lesssim 50$~keV), the small number density of SM particles the DM can scatter off imposes that the DM coupling has to be large for the DM to remain in kinetic equilibrium with the SM particles. Such a DM coupling large enough to realize $T_{kd} \lesssim m_e/10$  is however severely constrained by collider and DM search experiments, as we discuss in \S \ref{secexpbounds}.   




\begin{figure}
\begin{center}    
\epsfxsize = 0.48\textwidth
\includegraphics{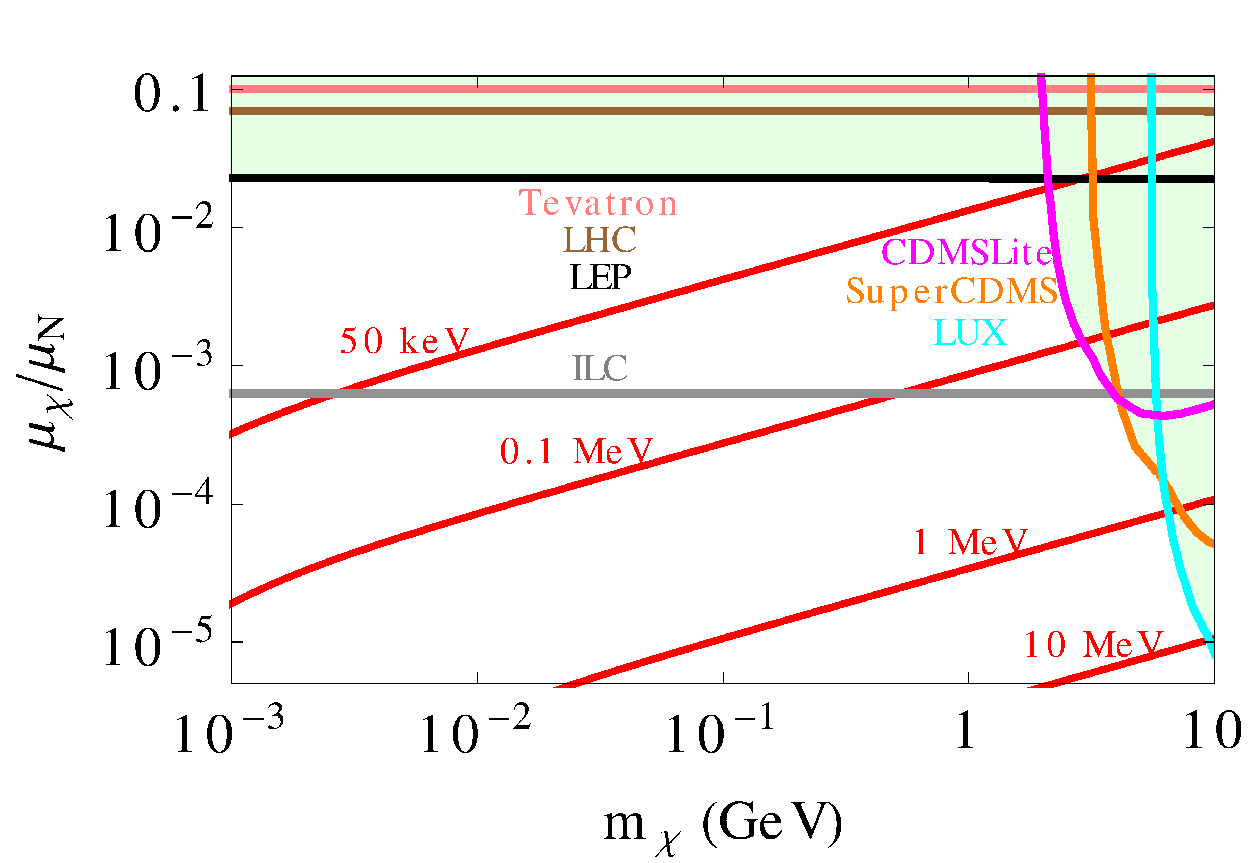}
\end{center}  
\caption{Experimental bounds on the DM magnetic dipole from collider and direct DM search experiments. The regions above the curves are excluded, and the currently excluded region is shaded with light green color. For the future prospects on the bounds for the light DM, the ILC bound is shown as a reference. The red contours show the DM kinetic decoupling temperature.} 
\label{Sept25temp}
\end{figure}


\section{Protohalo mass}
\label{secproto}

As an application of our dipole DM kinetic decoupling study, we estimate the smallest protohalo size for a given dipole moment $\mu_{\chi}$ and mass $m_{\chi}$.

The smallest protohalo mass can be estimated from the horizon size at kinetic decoupling and the free streaming scale of DM once decoupled from the thermal bath \cite{matias,bert}.
The horizon size is relevant here to quantify the protohalo mass because acoustic oscillations due to the coupling between DM and SM fluid damp the DM fluctuations inside the horizon. The mass inside the horizon at kinetic decoupling is
\ba
\label{macou}
M_{kd}\approx  \frac{4 \pi}{3} \frac{\rho_m(T_{kd})}{H^3(T_{kd})}, 
\ea
where $\rho_m$ is the matter density.
 DM free streaming is also relevant for the estimation of the protohalo mass because free streaming causes the erasure of fluctuations on scales smaller than the free-streaming length
\ba
\lambda_{fs}=a(t_0)\int^{t_0}_{t_{kd}} \frac{v(t)}{a(t)} \, dt.
\ea
Here $a(t)$ is the universe scale factor and $v(t)$ is the DM particle velocity.
The free streaming length $\lambda_{fs}$ is the distance a DM particle can travel without interacting with the plasma from the time of kinetic decoupling $t_{kd}$ to the present time $t_0$. DM fluctuation modes with wave number greater than $k_{fs}=2\pi/\lambda_{fs}$ are suppressed, and the protohalo cutoff mass from  free streaming effects can be estimated as the DM mass contained inside a sphere of radius $\pi/k_{fs}=\lambda_{fs}/2$,
\ba
\label{mfs}
M_{fs}\approx \frac{4 \pi}{3} \left(  \frac{\pi}{k_{fs}} \right)^3 \rho_{m}(t_0).
\ea
The smallest protohalo mass can be estimated by the greatest of $M_{kd}$ and $M_{fs}$, taking account of these two independent effects (acoustic oscillation damping and free streaming). The smallest possible protohalo mass can hence be obtained once the kinetic decoupling temperature and the DM mass are specified. 

Our results for the minimum protohalo mass are shown  in Fig. \ref{massSept25temp} as a function of $m_\chi$ and $\mu_\chi$. We obtain masses as large as $10^3~M_\Sun$, which, as found in the next section, are not excluded by current collider and DM direct search constraints. We note that for the conventional WIMPs such as those in supersymmetric models, the typical DM protohalo size is around the earth mass $10^{-6}M_{\Sun}$, and the effective operator analysis also typically bounds the protohalo mass to be less than the earth mass from collider and DM search experiments \cite{gon,pro4,kam2,wimpy,brifirst}. Such a $10^{-6}M_{\Sun}$ mass is far below the mass scale the current numerical simulation studying the large scale structures can resolve. Our light magnetic dipole DM scenarios allow for much larger protohalo masses.

\section{Collider and direct search experiment constraints}
\label{secexpbounds}
Constraints from collider and DM direct search experiments can be complementary to each other for light DM particles, in that such a particle can be kinematically accessible at colliders while direct search experiments lose sensitivity to light DM due to small recoil energies.
\subsection{Collider constraints}
Collider constraints on the DM magnetic dipole moment from LEP give tighter constraints than the Tevatron and the LHC \cite{tim11,barg12}. We show these constraints in Fig \ref{Sept25temp}. The light DM masses of interest to us are within the kinematic reach of the colliders, and the lepton collider has an advantage over the hadron collider because of its smaller background. We refer the reader to \cite{tim11} for more details on the comparison of the LEP, Tevatron and LHC bounds shown in Fig.~\ref{Sept25temp}. 

We add to Fig.~\ref{Sept25temp} the future reach of the ILC regarding the kinetic decoupling of magnetic dipole DM \cite{brau1,brau2,abe1,howie}. We show constraints from the ILC mono-photon signals $e^+ e^- \rightarrow \nu \bar{\nu} \gamma$ assuming a center of mass energy $\sqrt{s}=$ 1 TeV and an integrated luminosity 500 fb$^{-1}$. For the background estimation, we simply consider the main irreducible background $e^+ e^- \rightarrow \nu \bar{\nu} \gamma$.  In our analysis, the polarizations of electron and positron beams are assumed to be $(P_{-},P_+)=(0.8,0.5)$, which helps in reducing the backgrounds from the $t$ and $u$ channel W bosons exchange due to its $V$-$A$ couplings. 
We also applied the selection cuts $E_{\gamma}>8$~GeV and $|\cos \theta_{\gamma}|<0.995$, accounting for the detector loss of acceptance of events that are too close to the beam line or too soft. This also helps in further reducing the co-linear and infrared-enhanced backgrounds.
A cut $E_{\gamma} \leq 450$~GeV was also applied to further suppress the $s$-channel on-shell Z recoil contributions. 
For a sufficiently large number of events, the event probabilities become Gaussian and we put upper bounds on the dipole moment by requiring that the total number of background and signal events along with the statistical fluctuations fit in the 90\% confidence interval of the background events. For the light DM mass range of our interest ($m_\chi \lesssim O(10)$ GeV), the increase of the collider energy does not improve the bounds significantly, and the improvement of the bounds mainly can come from the increase in the luminosity.
We combine these collider constraints with those from the DM direct search experiments in the following subsection.

\subsection{Direct dark matter search constraints}

For the constraints from the DM direct search experiments, the differential cross section for scattering of the magnetic dipole DM with a target nucleus is (this is the non-relativistic limit of Eq. (\ref{relaform}))  \cite{barg2,indep,del}
\ba
\frac{d \sigma}{dE_R}
=
\mu_{\chi}^2 \frac{\alpha}{2} \frac{ m_T}{m_{\chi T}^2}\frac{1}{v^2}
\left[
Z^2 \left(\frac{v^2}{v_{min}^2} -1+\frac{m_{\chi T}^2}{m_{\chi}^2}\right)
  F_{E,T}^2({\bf q}^2)
  +
2\frac{\mu_T^2}{\mu_N^2}\frac{m_{\chi T}^2}{m_N^2}\left( \frac{J_T+1}{3J_T}\right)  F_{M,T}^2({\bf q}^2)
\right].
\label{dsigma}
\ea
Here $m_T$ is a target nucleus mass, $m_N$ is the nucleon mass, and $m_{\chi T}$ is the DM-target nucleus reduced mass. Furthermore, $\mu_T$ is the nucleus magnetic moment and $\mu_N$ is the nuclear magneton.
The first term in Eq.~(\ref{dsigma}) represents the coupling of the DM magnetic dipole to the nuclear electric field and the second term corresponds to the coupling of the DM magnetic dipole and the nuclear magnetic field.

The constraints on the DM magnetic dipole moment from direct DM search experiments have been studied in \cite{del}. In Fig.~\ref{Sept25temp} we show the 90\% confidence level upper bounds on $\mu_{\chi}$ from CDMSLite, SuperCDMS and LUX \cite{cdmslite1,cdmslite2,agne,lux1,lux2} (see \cite{del} for details). 
We extend the analysis in \cite{del} to  the lower DM mass region in Fig. \ref{Sept25temp} by applying the Maximum Gap Method \cite{yel} to the CDMSLite data \cite{cdmslite1}. For this, we use the Helm form factor for $F_E$ with the normalization $F_E(0)=1$ \cite{smith1,smith2,gondoloFF} while we ignore the momentum dependence in $F_M$ for simplicity. The DM-nucleus scattering rate (in counts/kg/day/keV) is
\ba
\frac{dR}{dE_R}=\frac{\rho_{\chi}}{m_{\chi}m_T}\int _{v\ge v_{min}(E_R)} 
d^3 v \, f(v,t) \, v \, \frac{d \sigma_T}{d E_R}(E_R,v),
\ea
where $\rho_{\chi}$ is the local DM mass density. 
The function $f(v,t)$ is the DM velocity distribution in the earth frame. 
We use the conventional Maxwellian distribution, truncated at a maximum escape velocity $v_{esc}$ (in the  rest frame of the galaxy), $f(\vec{v}_{\chi})\propto \exp ({-(\vec{v}_{\chi}+\vec{v}_E)^2}/{v_0^2})   \, \theta(v_{esc}-|\vec{v}_{\chi}+\vec{v}_E|)
$
with the normalization $\int _{v<v_{esc}} d^3 v_{\chi} f(\vec{v}) =1$ \cite{maxwell}. We used $\rho_{\chi} = 0.3$~GeV/cm$^3$, $v_0 = 220km/s$, $v_{esc} = 544$~km/s and the earth velocity in the galactic frame $v_E=232$~km/s.
The event rate as a function of the measured energy $E'$ is
\ba
\frac{dR}{dE'}=\epsilon(E') \int^{\infty}_0 dE_R \, G (E_R,E')\, \frac{dR}{dE_R},
\ea
where $\epsilon$ is the counting efficiency, $G(E_R,E')$ is the resolution function representing the probability for a recoil energy $E_R$ to be measured as $E'$. For this we calculated the mean value $E'$ by taking account of the quenching factor $Q(E_R)=E'/E_R$ and assuming the detector resolution to be Gaussian. Our analysis uses the energy range from 0.17 to 7.00 keVee, and takes
the efficiency $\epsilon(E')=0.985$ and the energy resolution $\sigma=14$ eVee.
We use the quenching factor \cite{cdmslite1,cdmslite2} 
\ba
E'=Q(E_R)E_R=\frac{1+Y(E_R)eV_b/\epsilon_{\gamma}}{1+e V_b/\epsilon_{\gamma}} E_R
\ea
with a voltage bias $V_b=69$~V and the average excitation energy to create an electron-hole pair $\epsilon_{\gamma}=3$~eV. The ionization yield is, for a nucleus of $Z$ protons and atomic mass $A$ \cite{smith2},
\ba
Y(E_R)=k\frac{g(\epsilon)}{1+kg(\epsilon)}
\ea
with $g(\epsilon)=3 \epsilon^{0.15}+0.7 \epsilon^{0.6}+\epsilon, \epsilon=11.5 E_R Z^{-7/3}$ and $k=0.133 Z^{2/3}A^{-1/2}$.

The bounds from the direct DM search experiments become
$\mu_{\chi}/\mu_N \lesssim 10^{-5},  2 \times 10^{-4}, 10^{-3}$ for $\mchi=10, 5, 3$ GeV, respectively. The current direct search experiments are not sensitive to magnetic dipole DM with $\mchi \lesssim 2$ GeV, as apparent in Fig. \ref{Sept25temp}. We can see that, for magnetic dipole DM with $m_{\chi}\gtrsim 5$ GeV to be consistent with the current direct search experiments,  kinetic decoupling should have occurred before the beginning of the electron-positron annihilation epoch ($T\sim m_e$). We can also find that the current collider and direct search experiments require that magnetic dipole DM with $m_{\chi} \gtrsim 2$ GeV should decouple kinetically before the electron-positron annihilation ceases to be efficient ($T_{kd} \gtrsim m_e/10 \sim 50$ keV). For $m_{\chi} \gtrsim 3$ MeV, a null search at the ILC would force $T_{kd}\gtrsim m_e/10$. For an even lighter mass, DM with sufficiently large magnetic dipole moment $\mu_{\chi}$ can stay in kinetic equilibrium even after the electron-positron annihilation epoch ($T_{kd} \lesssim m_e/10$), even though, due to the small number density of baryons and electrons the DM can elastically scatter off, $T_{kd}$ far below $m_e/10$ would require a large DM coupling strength which is severely constrained by colliders.

\section{Discussion/conclusion}
\label{secdis}

There is a further bound (a cosmological bound) on magnetic dipole DM that is however much weaker than those presented above. It arises from the fact that DM elastic scattering with baryons can cause the suppression of the DM fluctuation growth in the early Universe. This suppression comes from the dragging of DM fluctuations by the baryon fluid, and it constrains  the transport cross section $\sigma_T$ \cite{cora,celine,sig,chen2,boesh}. Ref. \cite{cora}, for instance, obtained the constraint $\sigma_T/m_{\chi}\lesssim 3.3\times 10^{-3} cm^2/g$ from the Planck CMB and SDSS Lyman-$\alpha$ forest data \cite{pla,lym}, which leads to the requirement 
\ba
\left(
\frac{\mu_{\chi}}{\mu_N}
\right)^2
\lesssim
7.5 \cdot 10^4 
\left(
1+
\frac{1+2(m_{\chi}/m_p)^2}{\left(1+(m_{\chi}/m_p)\right)^2}
\right)^{-1}
\left(
\frac{m_{\chi}}{m_p} 
\right).
\ea
This cosmological bound was obtained for a DM mass larger than the proton mass \cite{cora}, and hence, for a DM mass to which the current direct DM search experiments are sensitive, the cosmological bound from DM fluctuation suppressions due to momentum transfer in DM elastic scattering is currently much weaker than the other bounds we have discussed in this paper. The quantitative analysis of the constraints from the large scale structure in the Universe in the limit of $m_{\chi}<m_p$ is beyond the scope of our paper and we leave it for our future work to be compared with our collider bounds.

We found the current direct DM search and collider experiments allow light magnetic dipole DM to remain in  kinetic equilibrium as late as the electron-positron annihilation epoch. The allowed protohalo mass turns out to be ${\cal O}(10^{3})M_{\Sun}$ for $m_{\chi}\sim 1$ GeV, and larger protohalo masses are enforced for lighter DM. This is in contrast to the typically quoted protohalo masses of order the earth mass ${\cal O}(10^{-6})M_{\Sun}$ for the conventional weak scale DM \cite{gon,pro4}. The clarification of subhalo size clumpiness would be of great interest for the interpretation of DM search experiments which rely on the assumption of a smooth DM density \cite{blanc,bere,oda,cyr}.

Even though we have here focused on studies only down to the electron-positron annihilation epoch, further analysis of an even lower kinetic decoupling temperature would be worth exploring. 
We plan to come back to such issues regarding a very low DM kinetic decoupling temperature beyond the range covered in this paper in more general scenarios, including other experimental bounds such as  indirect DM search experiments. 

\section*{Acknowledgment}
PG was supported by NSF (PHY-1415974) and KK was supported by Institute for Basic Science (IBS-R018-D1). We thank J. Huh for the useful discussions and PG thanks the CTPU at IBS for hospitality.

\end{document}